\documentclass{article}

\usepackage{arxiv}
\usepackage[utf8]{inputenc}
\usepackage[T1]{fontenc}
\usepackage{hyperref}
\usepackage{url}
\usepackage{booktabs}
\usepackage{amsfonts}
\usepackage{nicefrac}
\usepackage{microtype}
\usepackage{graphicx}
\usepackage{natbib}
\usepackage{doi}
\usepackage{caption}
\usepackage{subcaption}
\usepackage{algorithm}
\usepackage{algorithmic}
\usepackage{amsmath}

\title{An Evolutionary Hypothesis for White’s Illusion: \\ 
Constant Percentage Correction and Modern Individual Variation}

\author{
\href{https://orcid.org/0000-0002-9726-1340}{\includegraphics[scale=0.06]{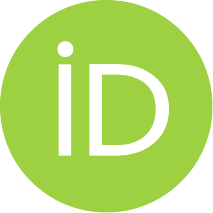}\hspace{1mm}Bekir Tevfik Akgün} \\
Faculty of Computer and Information Sciences \\
Yeditepe University, Turkey \\
\texttt{bekirtevfik.akgun@yeditepe.edu.tr}
\and
\href{https://orcid.org/0000-0003-3618-4166}{\includegraphics[scale=0.06]{orcid.pdf}\hspace{1mm}Ergun Akleman} \\
Visual Computing \& Computational Media, SPVFA \\
Computer Science and Engineering, COE \\
Texas A\&M University, USA \\
\texttt{ergun@tamu.edu}
}

\begin{document}

\maketitle

\begin{abstract}
White’s illusion is a striking demonstration of context-dependent brightness and lightness perception.
In earlier work, we developed a forward translucency-based formula capable of producing
approximately constant perceived appearance across varying backgrounds. While validating that formulation,
we observed a notable population-level pattern: individuals selected slightly different
correction percentages for ``constant perception,'' while the population average remained
remarkably stable across experiments.

In this paper, we propose an evolutionary hypothesis for this phenomenon. We suggest that
a simple \emph{constant-percentage correction function} could have emerged as an adaptive
mechanism for maintaining useful material and reflectance judgments under variable
illumination. Under this interpretation, the stability of the population mean may reflect
a conserved perceptual mechanism, whereas individual differences may reflect learning, adaptation, or variation
around a shared computational response.

Using a modification of our previously published forward model, we show that White’s
illusion can be reproduced by applying a constant-percentage correction to background
luminance. The model therefore provides a compact computational account of the observed
population-level regularity and motivates a testable evolutionary interpretation of
individual variation in contextual lightness judgments.
\end{abstract}

\begin{figure}[hbtp]
\centering
\includegraphics[width=0.99\textwidth]{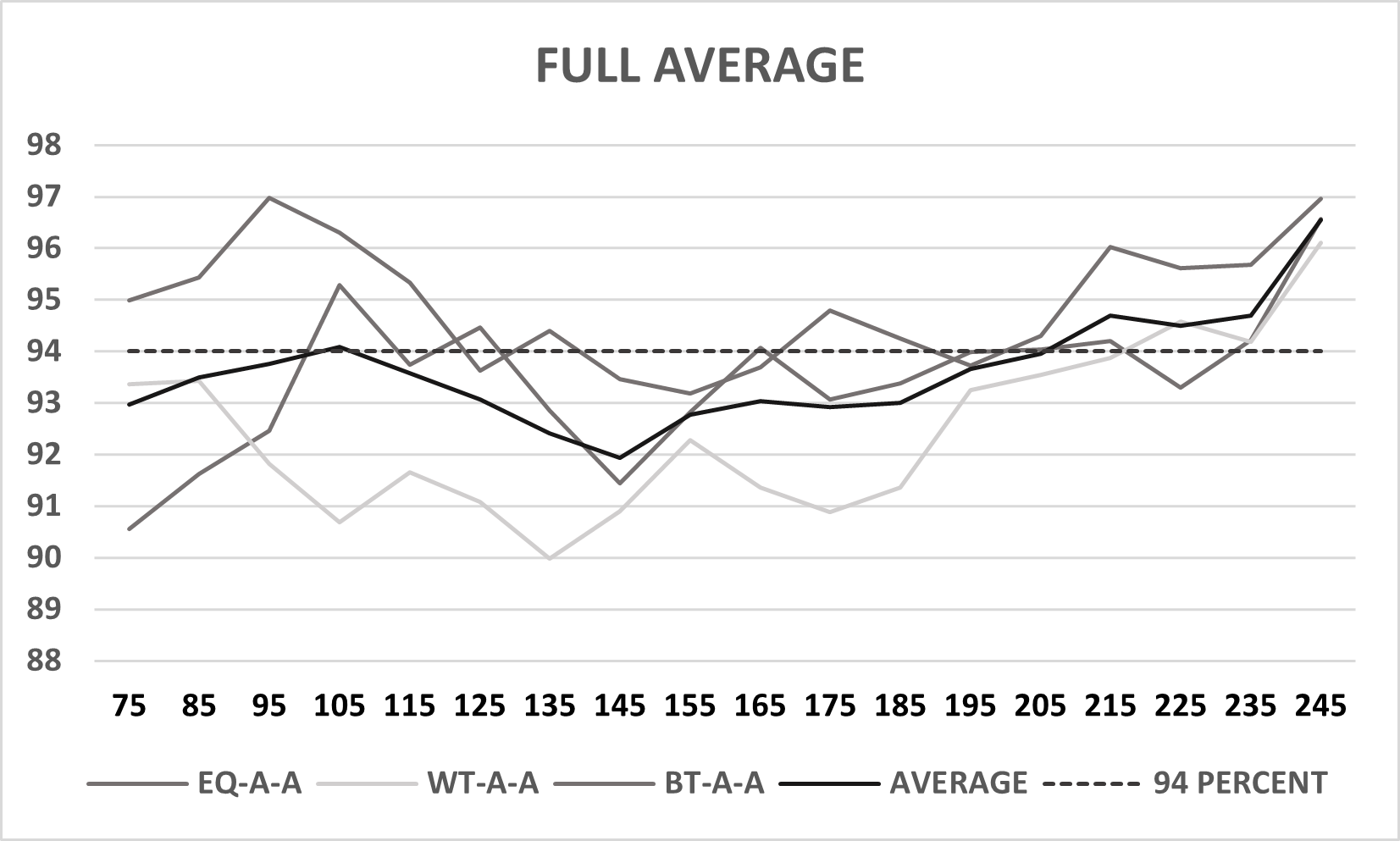}
 \hfill
    \caption{Illustration of the constant-percentage correction underlying the proposed model. The same correction percentage is applied across different local background luminances, providing the population-level reference behavior considered in this paper.}
\label{fig_flat}
\end{figure}

\begin{figure}[htpb]
    \centering        
        \begin{subfigure}[t]{0.48\textwidth}    \includegraphics[width=0.99\linewidth]{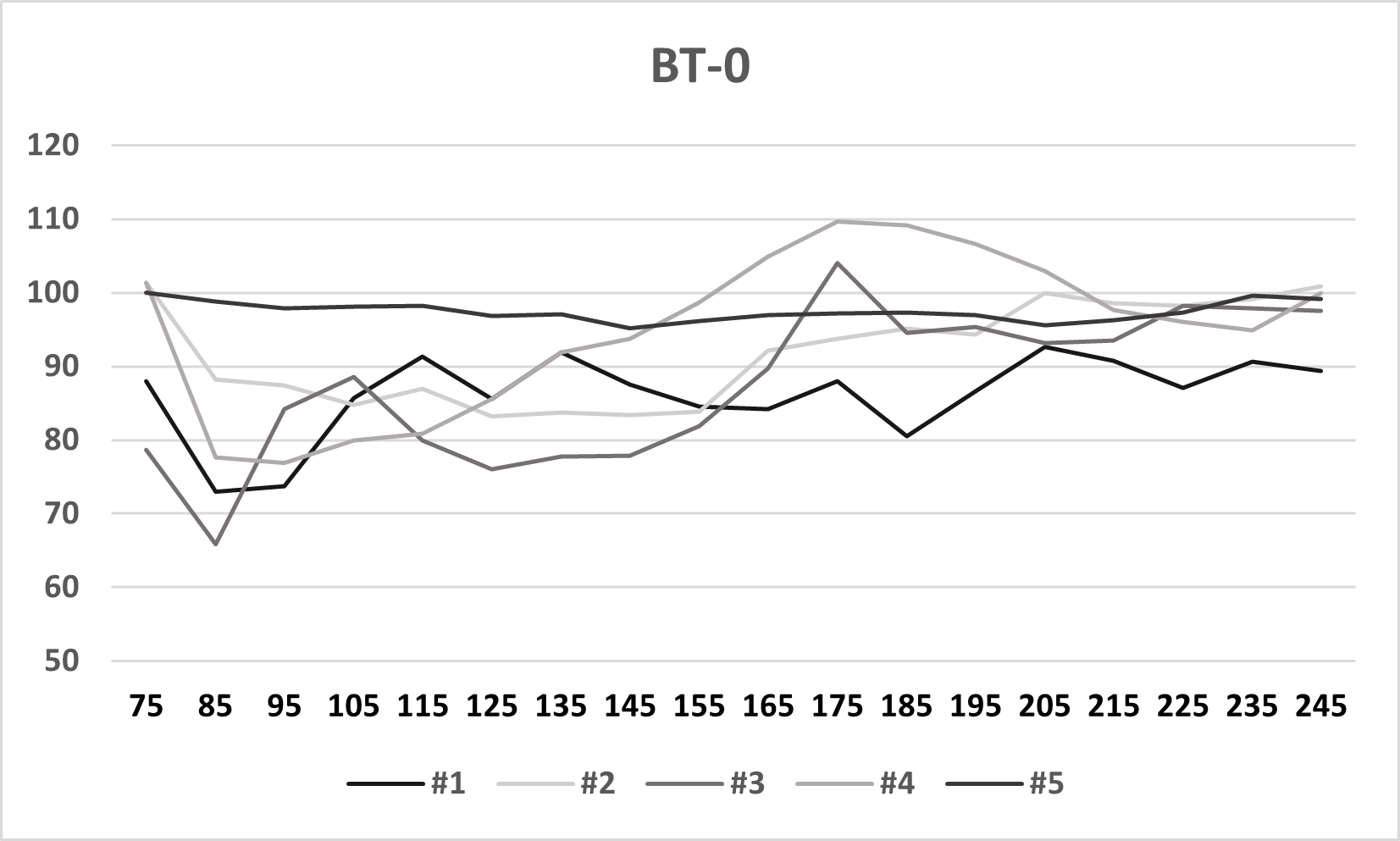}
            \caption{\it BT 000.}
        \label{fig_BT_000}
    \end{subfigure}
        \hfill
         \begin{subfigure}[t]{0.48\textwidth}    \includegraphics[width=0.99\linewidth]{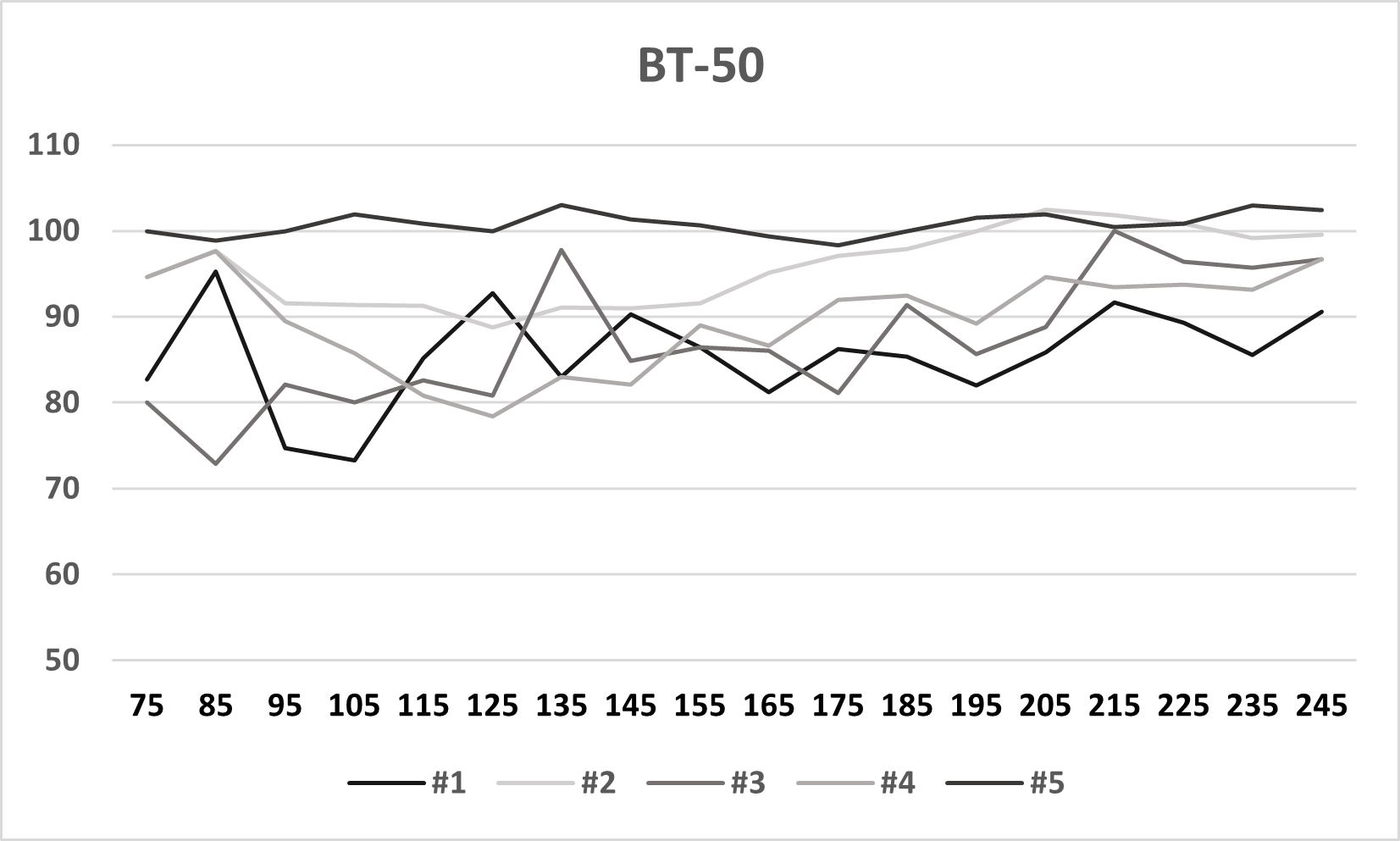}
            \caption{\it BT 050.}
        \label{fig_BT_050}
    \end{subfigure}
        \hfill
            \begin{subfigure}[t]{0.48\textwidth}    \includegraphics[width=0.99\linewidth]{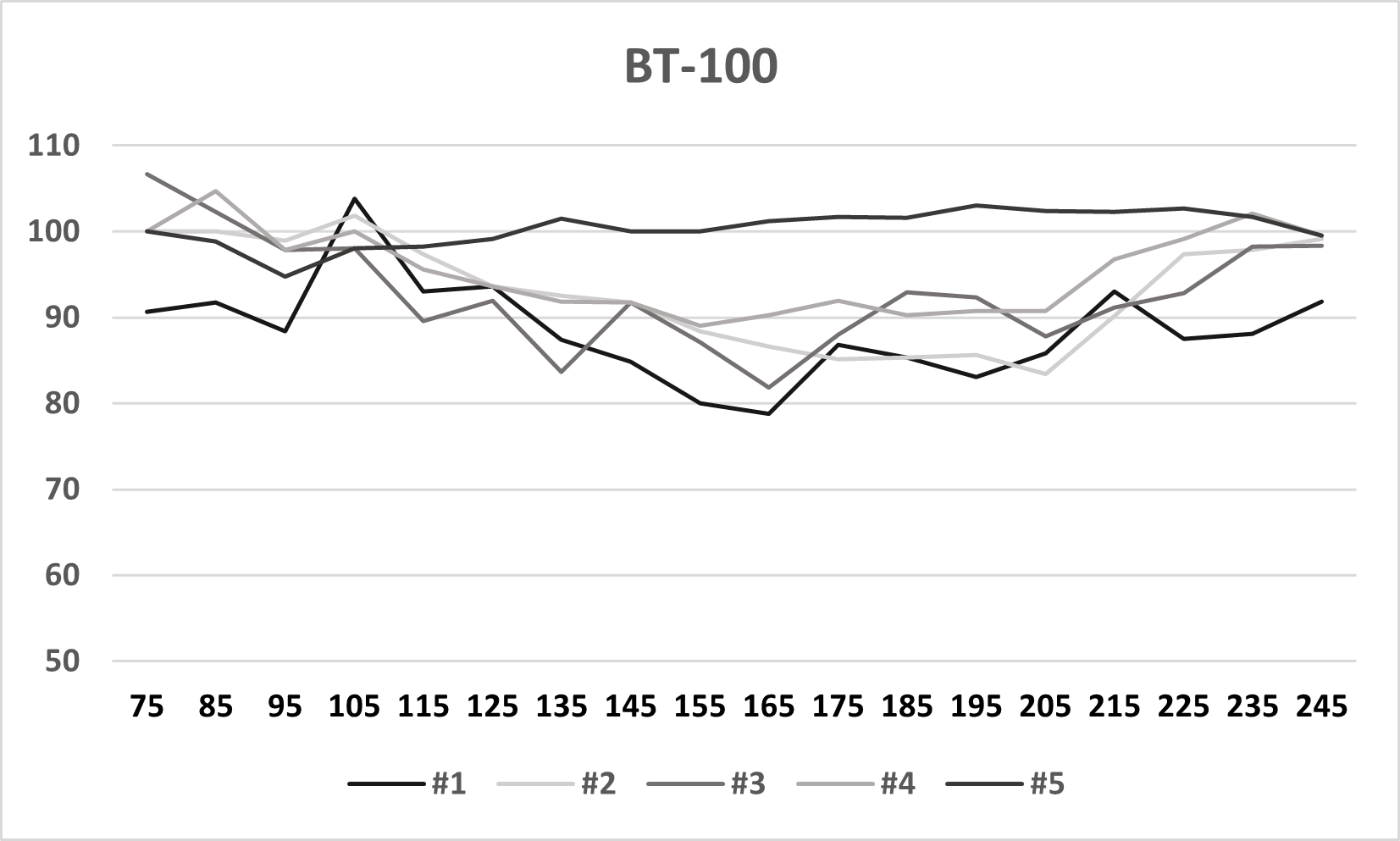}
            \caption{\it BT 100.}
        \label{fig_BT_100}
    \end{subfigure}
        \hfill
         \begin{subfigure}[t]{0.48\textwidth}    \includegraphics[width=0.99\linewidth]{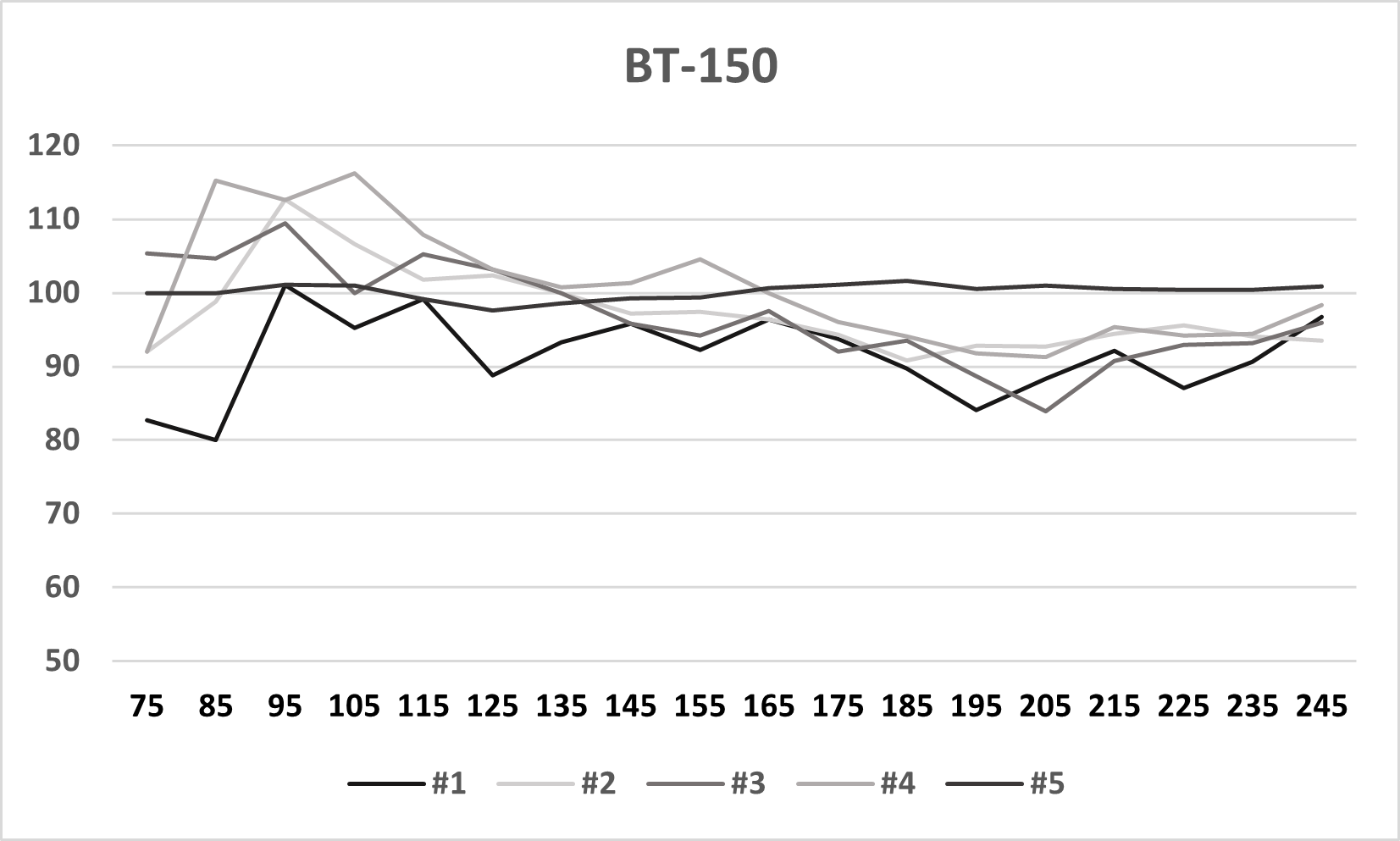}
            \caption{\it BT 150.}
        \label{fig_BT_150}
    \end{subfigure}
        \hspace{0.1in}
    \begin{subfigure}[t]{0.48\textwidth}    \includegraphics[width=0.99\linewidth]{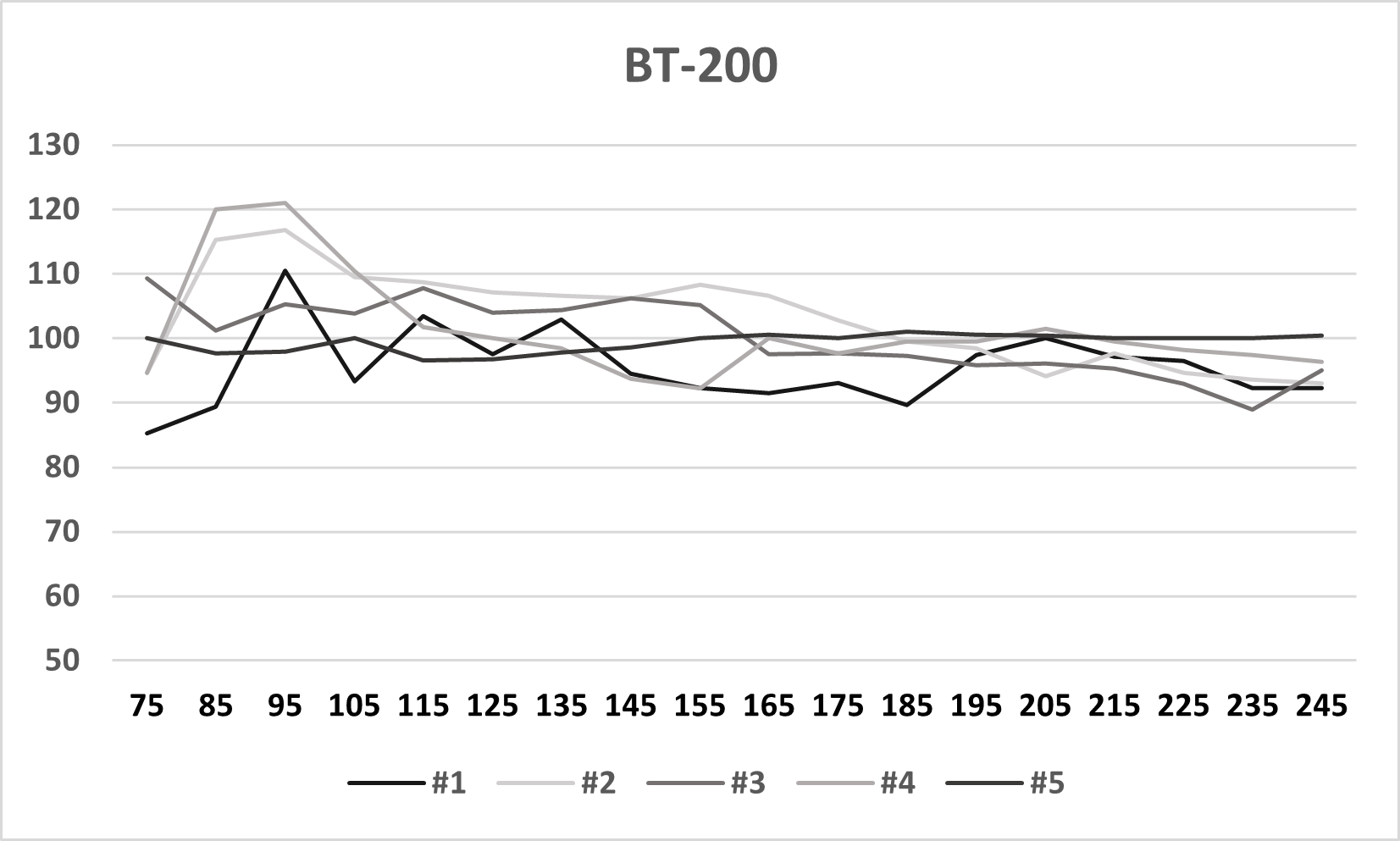}
            \caption{\it BT 200.}
        \label{fig_BT_200}
    \end{subfigure}
\caption{White's illusion stimuli in the black-thinner (BT) configuration, in which each black bar is narrower than each white bar. The panels retain their original stimulus identifiers.}
\label{fig_BT}
\end{figure}

\begin{figure}[htpb]
    \centering        
        \begin{subfigure}[t]{0.485\columnwidth}    \includegraphics[width=0.99\linewidth]{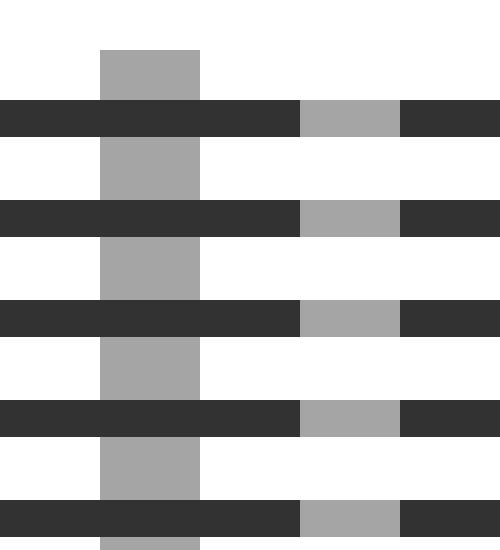}
            \caption{\it BT050165165.}
        \label{fig_BT_050_165_165}
    \end{subfigure} 
    \hfill
    \begin{subfigure}[t]{0.485\columnwidth}    \includegraphics[width=0.99\linewidth]{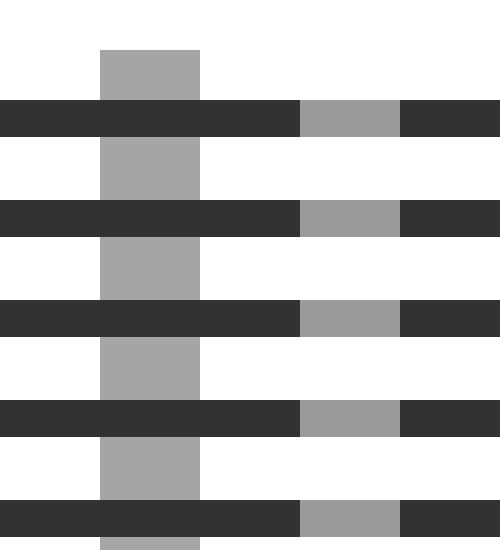}
            \caption{\it BT050165154.}
        \label{fig_BT_050_165_154}
    \end{subfigure}  
\caption{Representative stimuli for the black-thinner (BT) configuration, where black bars are narrower than white bars. The two panels show the original stimulus variants identified by their filenames.}
\label{fig_BT_Images}
\end{figure}

\begin{figure}[htpb]
    \centering        
        \begin{subfigure}[t]{0.48\textwidth}    \includegraphics[width=0.99\linewidth]{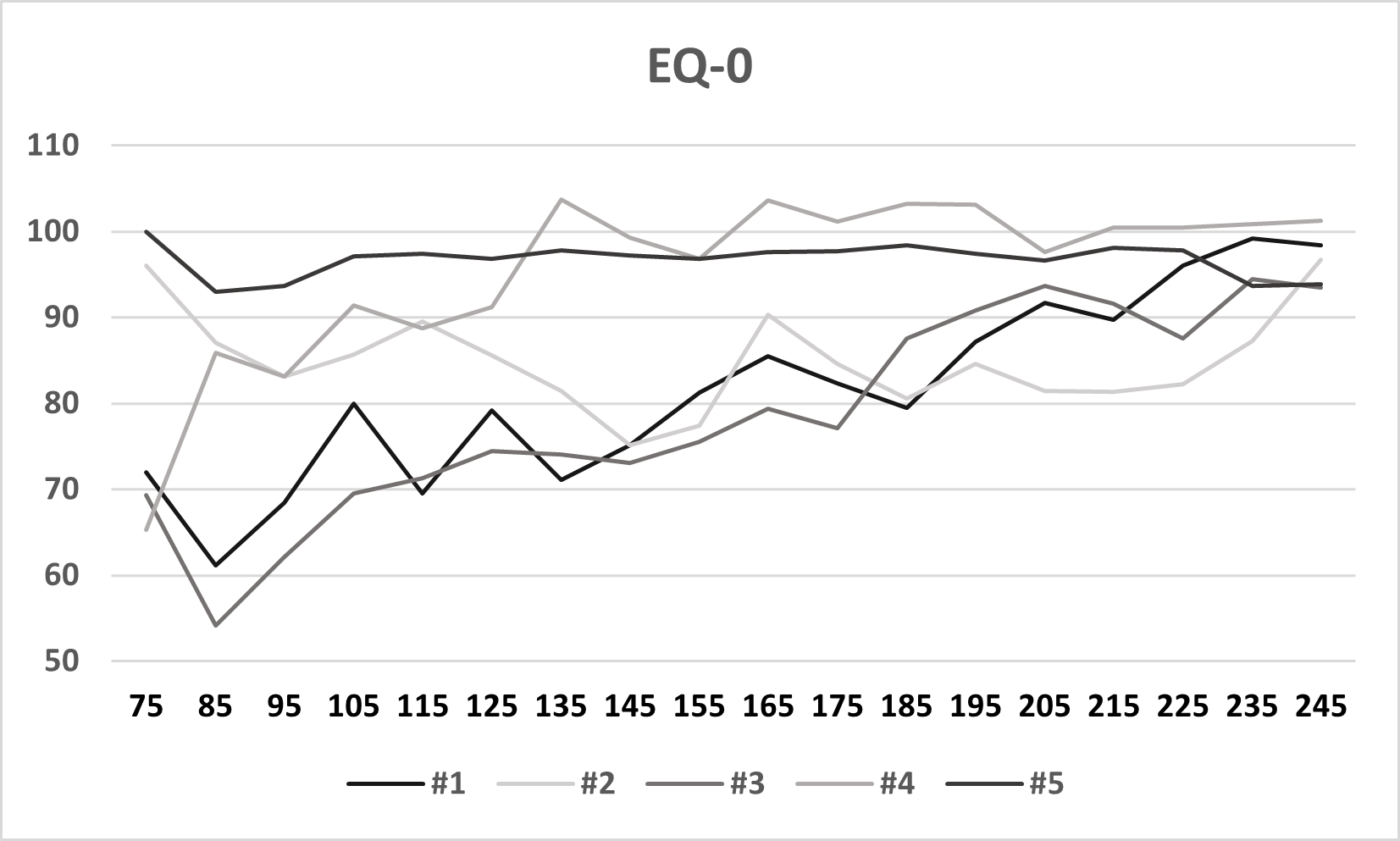}
            \caption{\it EQ 000.}
        \label{fig_EQ_000}
    \end{subfigure}
        \hfill
         \begin{subfigure}[t]{0.48\textwidth}    \includegraphics[width=0.99\linewidth]{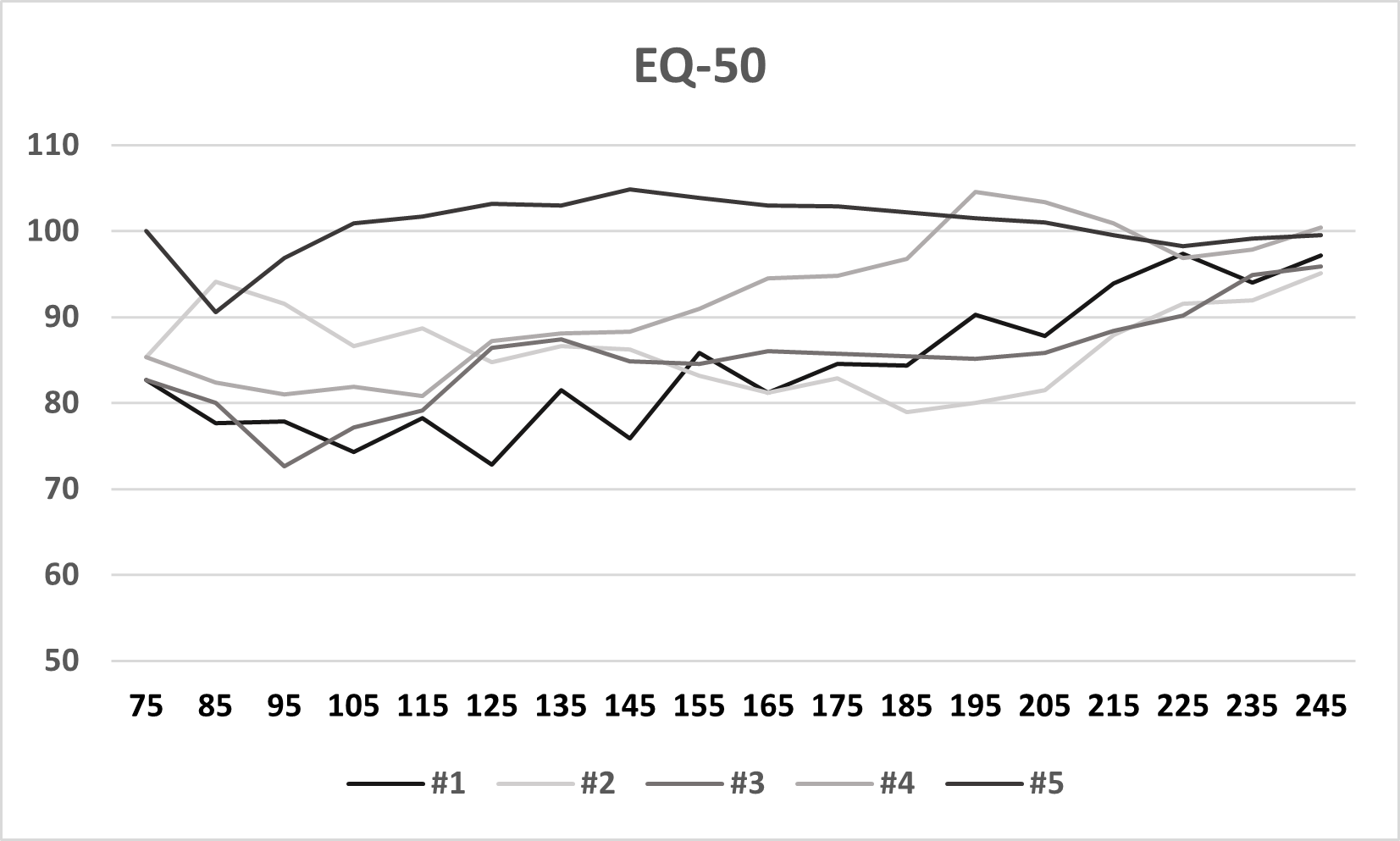}
            \caption{\it EQ 050.}
        \label{fig_EQ_050}
    \end{subfigure}
        \hfill
            \begin{subfigure}[t]{0.48\textwidth}    \includegraphics[width=0.99\linewidth]{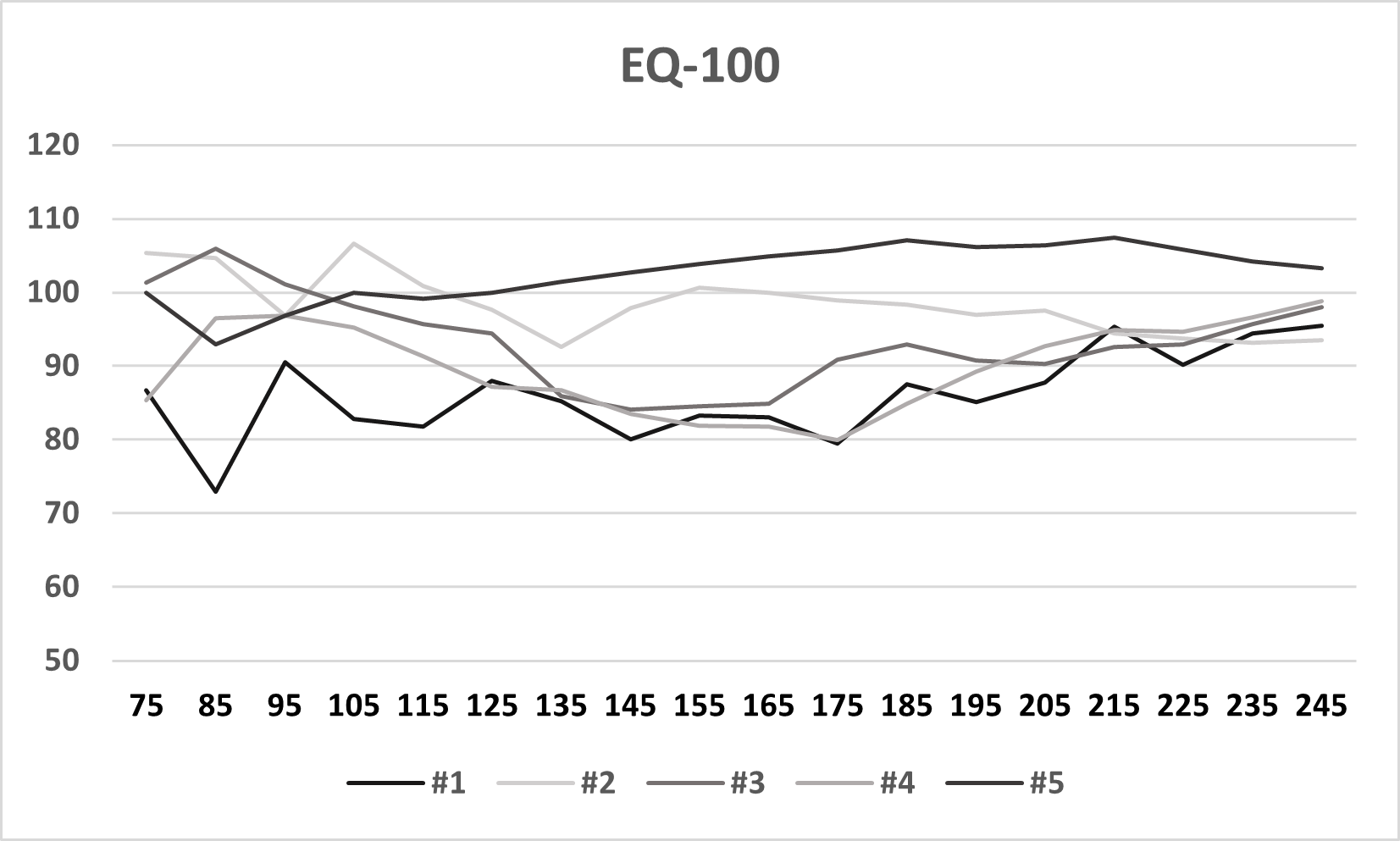}
            \caption{\it EQ 100.}
        \label{fig_EQ_100}
    \end{subfigure}
        \hfill
         \begin{subfigure}[t]{0.48\textwidth}    \includegraphics[width=0.99\linewidth]{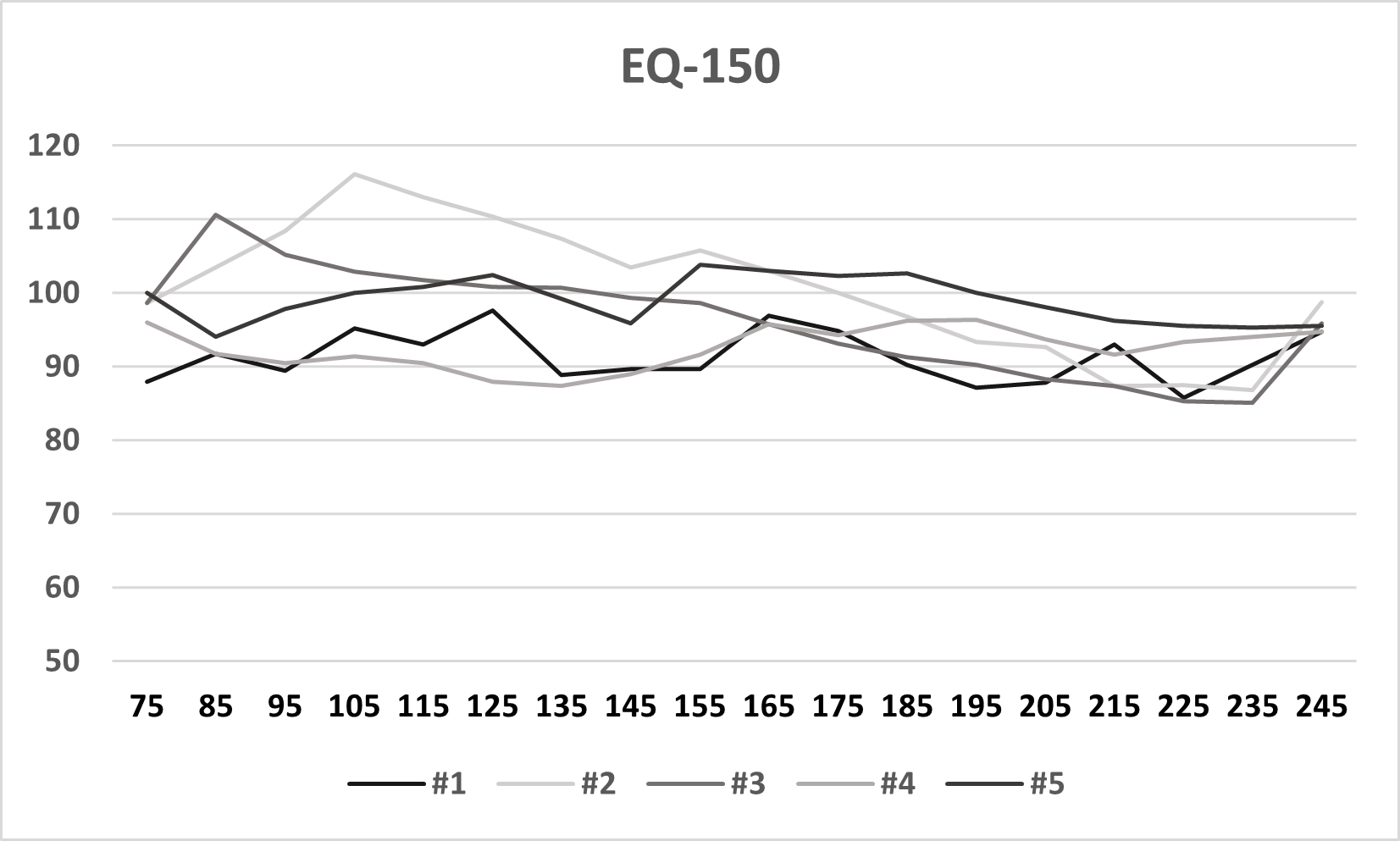}
            \caption{\it EQ 150.}
        \label{fig_EQ_150}
    \end{subfigure}
\hspace{0.1in}
    \begin{subfigure}[t]{0.48\textwidth}    \includegraphics[width=0.99\linewidth]{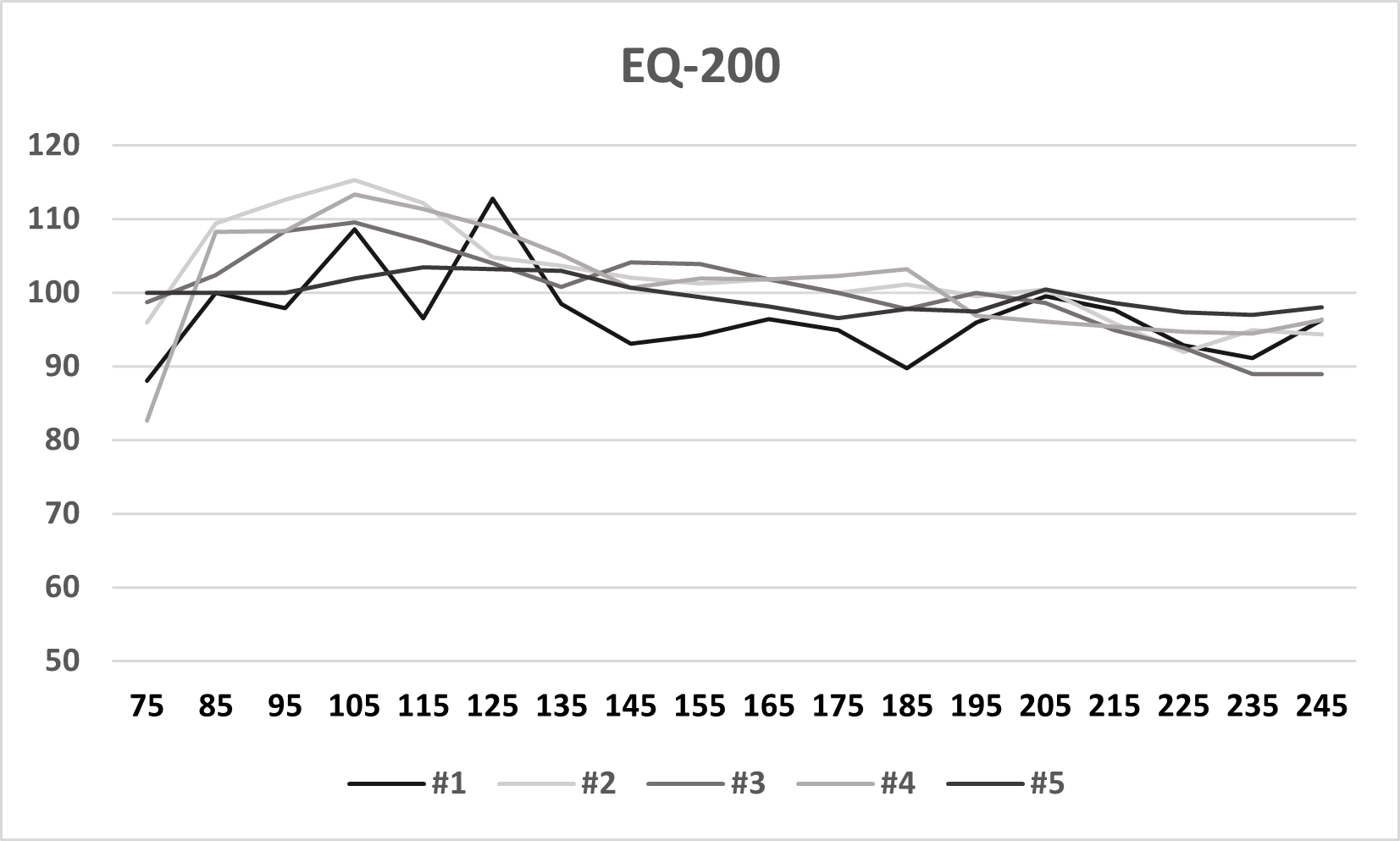}
            \caption{\it EQ 200.}
        \label{fig_EQ_200}
    \end{subfigure}
\caption{White's illusion stimuli in the equal-thickness (EQ) configuration, in which black and white bars have the same width. The panels retain their original stimulus identifiers.}
\label{fig_EQ}
\end{figure}

\begin{figure}[htpb]
    \centering        
        \begin{subfigure}[t]{0.485\columnwidth}    \includegraphics[width=0.99\linewidth]{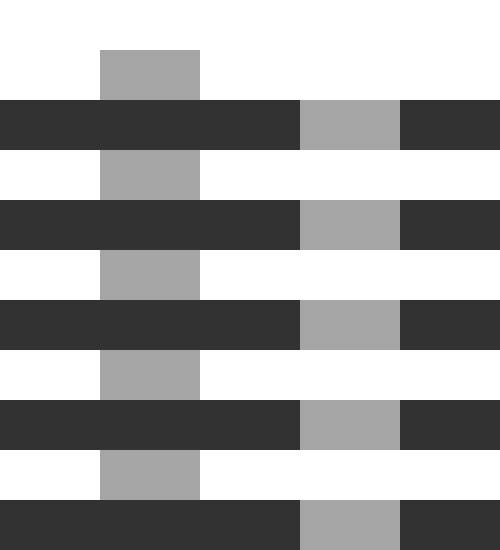}
            \caption{\it EQ050165165.}
        \label{fig_EQ_050_165_165}
    \end{subfigure} 
    \hfill
    \begin{subfigure}[t]{0.485\columnwidth}    \includegraphics[width=0.99\linewidth]{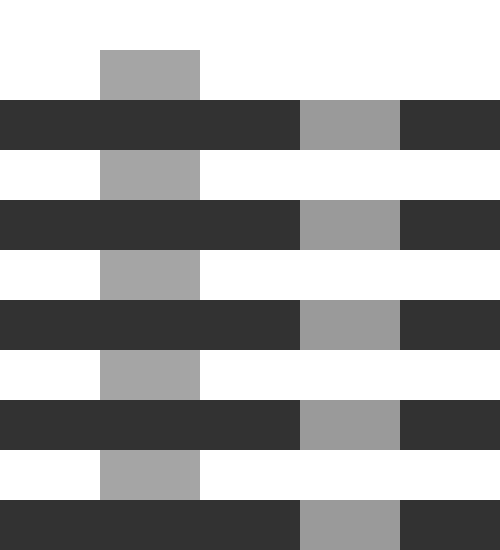}
            \caption{\it EQ050165154.}
        \label{fig_EQ_050_165_154}
    \end{subfigure}  
\caption{Representative stimuli for the equal-thickness (EQ) configuration, where black and white bars have identical widths. The two panels show the original stimulus variants identified by their filenames.}
\label{fig_EQ_Images}
\end{figure}

\section{Introduction}

White’s illusion, like many classical lightness illusions, reveals that perceived brightness and lightness depend strongly on context. In earlier work, we constructed a forward formula that produces a constant-appearance setting by blending a foreground patch with a blurred version of its background~\cite{akleman2024constant}. During that work, an empirical pattern emerged that motivates the present paper: although participants selected somewhat different correction percentages for ``perfect constancy,'' the population average produced a remarkably stable correction curve. As illustrated in Figure~\ref{fig_flat},
the population-level average remains approximately
constant across the tested background luminances,
despite variations among individual observers and
stimulus configurations.

The present study focuses on the interpretation of this population-level stability. Rather than treating individual differences as measurement noise around an arbitrary parameter, we ask whether the stable mean could reflect a simple perceptual mechanism with an evolutionary origin. Our proposal is intentionally framed as a hypothesis: the experiments establish individual variation together with a stable aggregate response, while the evolutionary account offers a possible explanation for that observation.

We therefore build on our earlier computational formulation and examine White’s illusion as a possible consequence of a potentially shared constant-percentage correction mechanism. The contribution of this paper is not a direct demonstration of evolutionary history, but a computationally explicit hypothesis that connects the observed stability of the mean response to a simple mechanism that could have been useful for visual judgments under variable illumination.

\begin{figure}[htpb]
    \centering        
        \begin{subfigure}[t]{0.48\textwidth}    \includegraphics[width=0.99\linewidth]{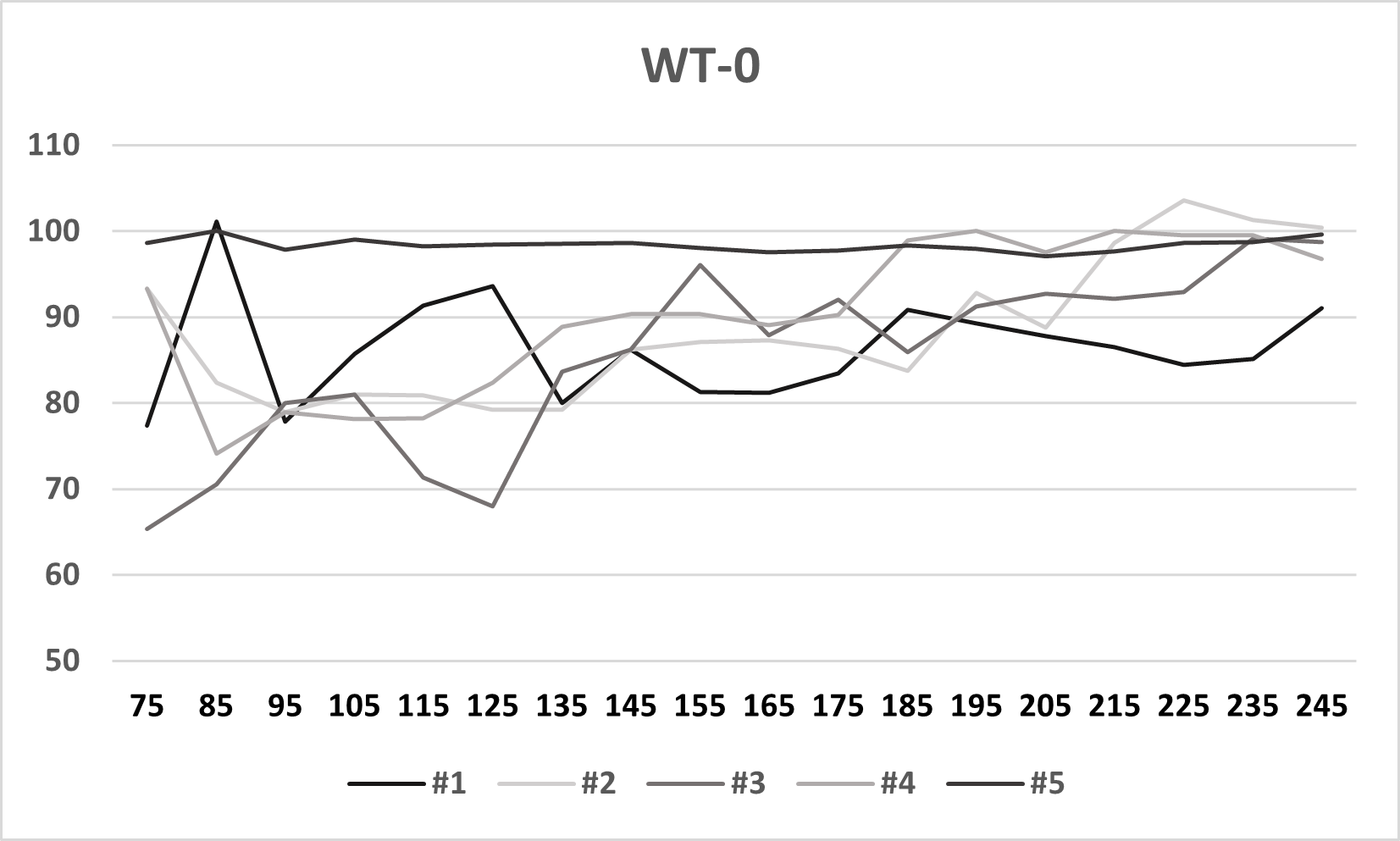}
            \caption{\it WT 000.}
        \label{fig_WT_000}
    \end{subfigure}
        \hfill
         \begin{subfigure}[t]{0.48\textwidth}    \includegraphics[width=0.99\linewidth]{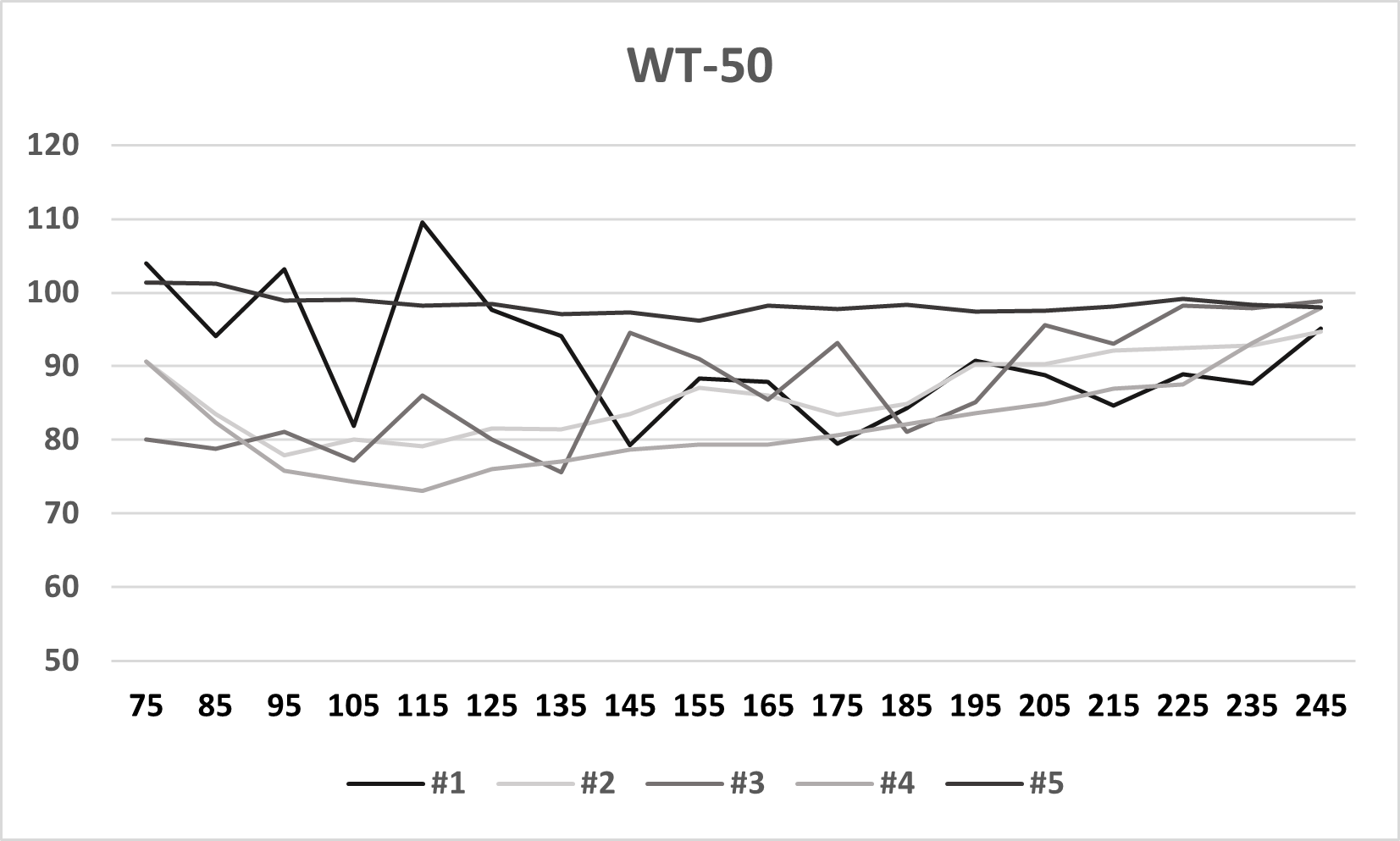}
            \caption{\it EQ 050.}
        \label{fig_WT_050}
    \end{subfigure}
        \hfill
            \begin{subfigure}[t]{0.48\textwidth}    \includegraphics[width=0.99\linewidth]{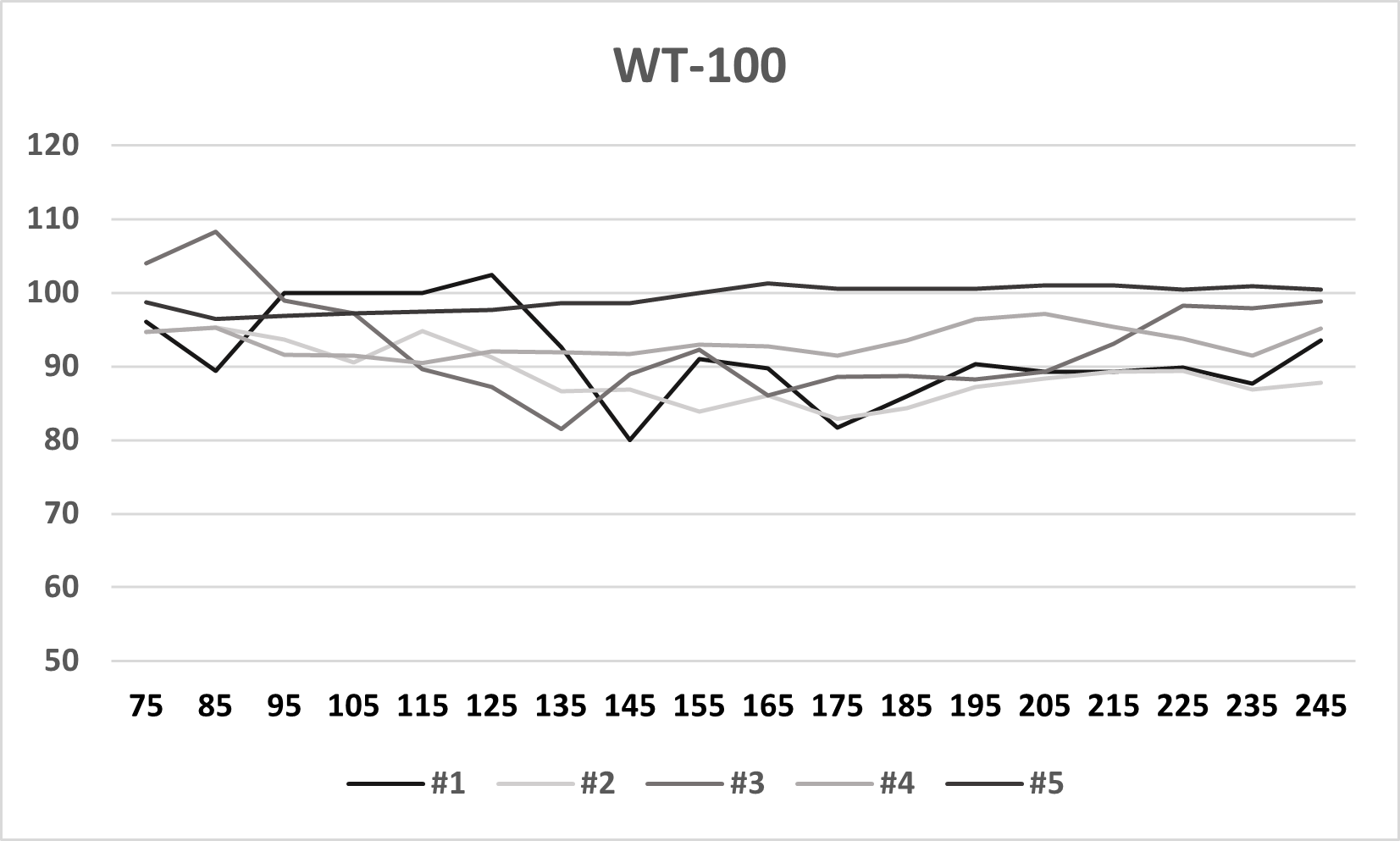}
            \caption{\it WT 100.}
        \label{fig_WT_100}
    \end{subfigure}
        \hfill
         \begin{subfigure}[t]{0.48\textwidth}    \includegraphics[width=0.99\linewidth]{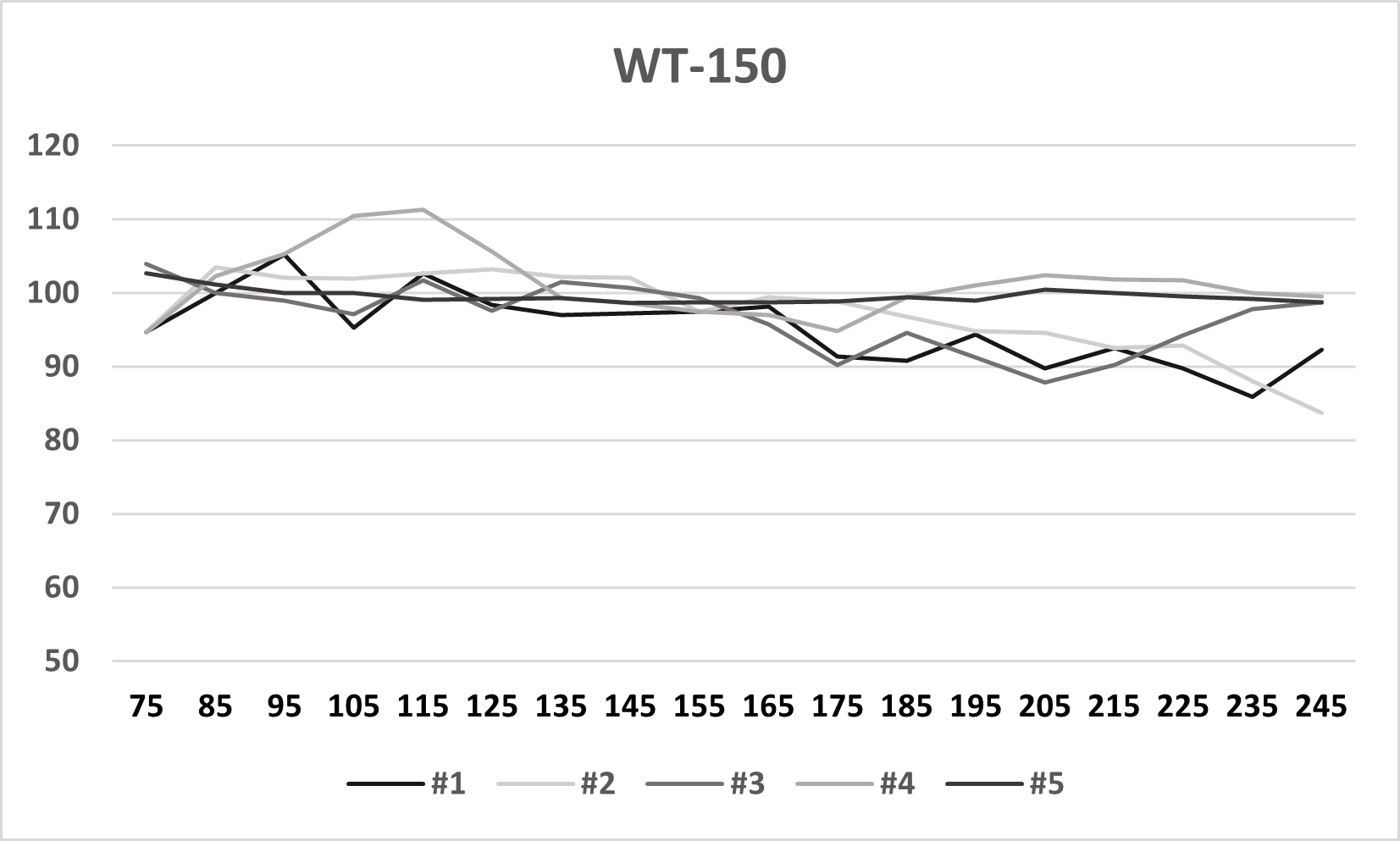}
            \caption{\it WT 150.}
        \label{fig_WT_150}
    \end{subfigure}
\hspace{0.1in}
    \begin{subfigure}[t]{0.48\textwidth}    \includegraphics[width=0.99\linewidth]{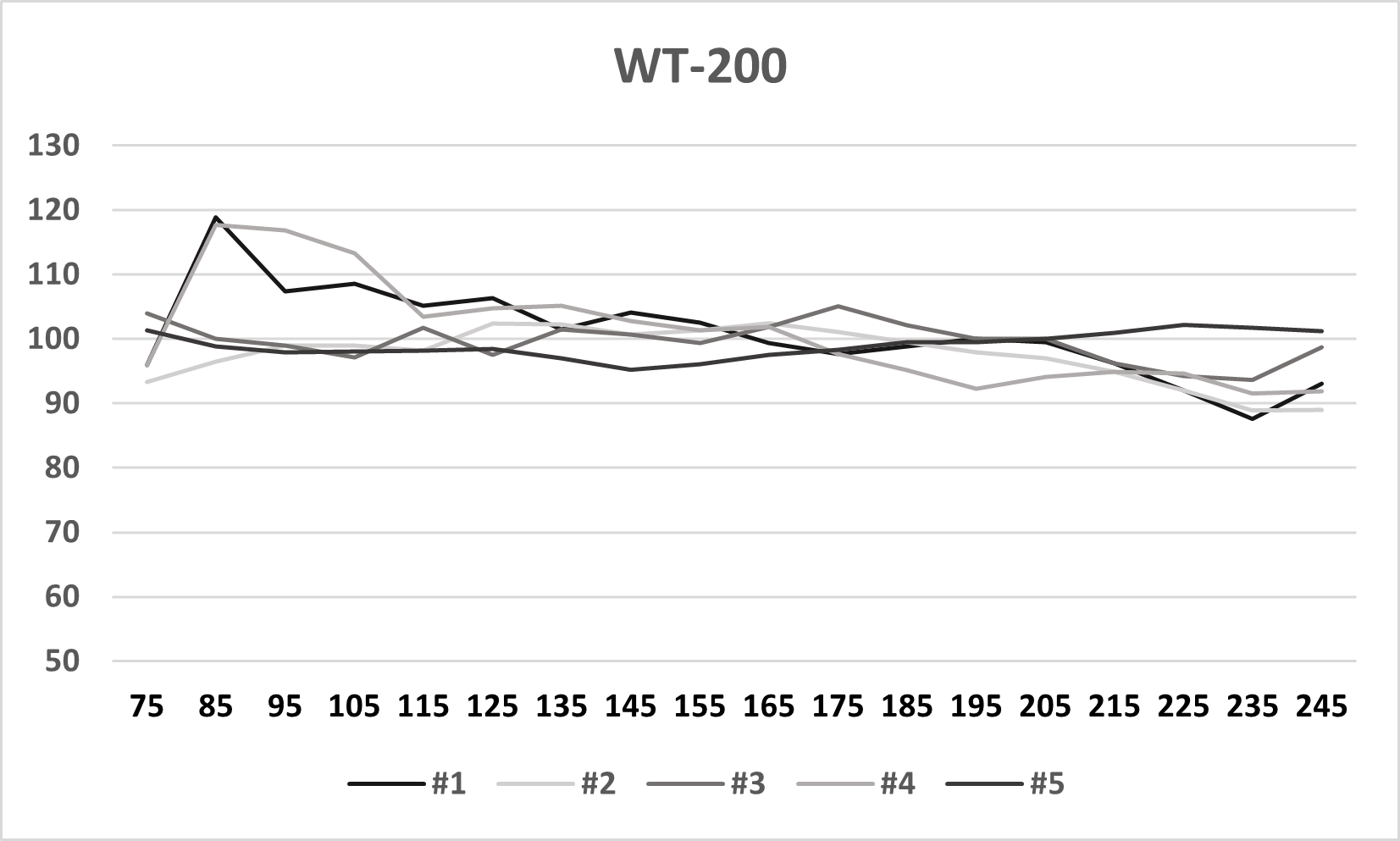}
            \caption{\it WT 200.}
        \label{fig_WT_200}
    \end{subfigure}
\caption{White's illusion stimuli in the thinner white (WT) configuration, in which white bars are thinner than black bars. The panels retain their original stimulus identifiers.}
\label{fig_WT}
\end{figure}

\section{Evolutionary Hypothesis}

Animals must recognize relevant surfaces, objects, food sources, and other organisms under shadows, occlusions, and changing illumination. A visual mechanism that reduced the effect of illumination on apparent surface properties could therefore have provided an adaptive advantage. We hypothesize that early visual systems may have developed a fast, low-complexity mechanism that approximated this compensation through a \emph{constant percentage correction} based on local average luminance.

Such a mechanism would not require an explicit physical reconstruction of illumination and reflectance. Instead, a fixed or slowly varying correction could provide a useful approximation across many natural viewing conditions. If such a mechanism became conserved, a population-level response could remain stable even when individuals exhibit modest variation in its effective parameter.

\begin{figure}[htpb]
    \centering        
        \begin{subfigure}[t]{0.485\columnwidth}    \includegraphics[width=0.99\linewidth]{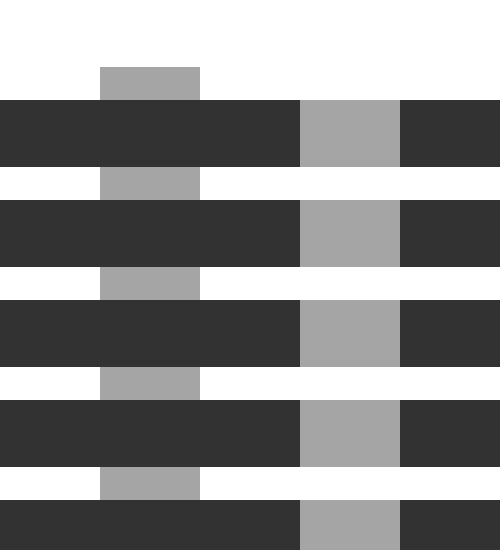}
            \caption{\it WT050165165.}
        \label{fig_WT_050_165_165}
    \end{subfigure} 
    \hfill
    \begin{subfigure}[t]{0.485\columnwidth}    \includegraphics[width=0.99\linewidth]{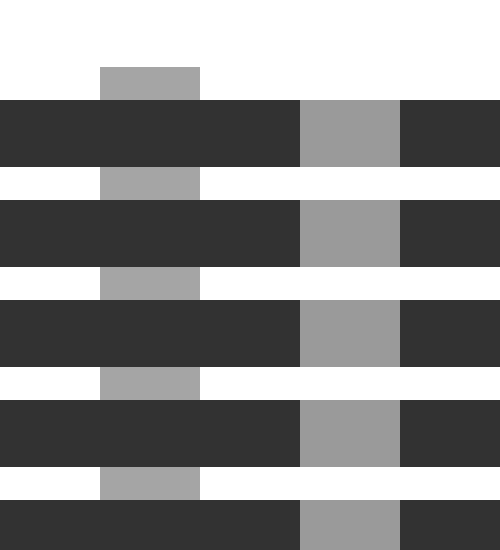}
            \caption{\it WT050165154.}
        \label{fig_WT_050_165_154}
    \end{subfigure}  
\caption{Representative stimuli in the thinner white (WT) configuration, in which white bars are thinner than black bars. The two panels show the original stimulus variants identified by their filenames.}
\label{fig_WT_Images}
\end{figure}

We further hypothesize that individual variation may arise from experience-dependent calibration, adaptive differences, or stochastic variation around a shared mechanism; these explanations cannot be distinguished with the current evidence. This interpretation is not established by the present experiments and should be regarded as a testable evolutionary explanation of the observed data. In particular, the present study does not identify a genetic mechanism or directly measure evolutionary change; it asks whether the combination of individual variation and a stable population mean is consistent with a conserved, low-complexity correction process.

\section{Related Work}

\paragraph{Terminology and scope.}
We use \emph{luminance} for a measurable photometric stimulus quantity, \emph{brightness} for perceived intensity, and \emph{lightness} for perceived surface reflectance or achromatic appearance. Our computational output is a stimulus luminance selected to elicit an approximately matched appearance; it is not a direct measurement of neural activity, inferred reflectance, or perceived luminance. The distinction matters because the experiments use perceptual judgments to calibrate a forward image-generation rule.

\paragraph{Established computational explanations of White's illusion.}
Oriented Difference-of-Gaussians (ODOG) filtering with normalization provides a prominent mechanistic explanation for White's effect and related brightness phenomena~\cite{blakeslee2016odog}. Variants discussed in the computational literature include locally normalized and frequency-dependent extensions. In a systematic comparison of eight computational models across thirteen displays with human observations, Nedimovi\'c and colleagues found that ODOG and several variants predicted White's illusion, while the FL-ODOG variant performed especially broadly across their stimulus battery~\cite{nedimovic2022evaluation}. These studies establish a substantial benchmark that a new descriptive model should not be claimed to surpass without direct quantitative evaluation. Our correction rule has a different and narrower goal: representing a family of appearance-matching settings with one interpretable mixing parameter, rather than predicting responses to arbitrary images from spatial filtering alone.

\paragraph{Ecological and empirical accounts.}
Corney and Lotto trained networks to infer surface reflectance in simulated three-dimensional scenes and reported that White-like assimilation emerged with three-dimensional rather than two-dimensional training environments~\cite{corney2007lightness}. Their result suggests that scene statistics and learned inference can generate perceptual effects without prescribing our specific blending rule. Purves and colleagues likewise argue that perceptual responses may reflect the empirical utility of prior experience rather than recovery of physical scene properties~\cite{purves2015perception}. These accounts are compatible with an economical approximate correction, but they do not establish that a constant percentage is innate or evolutionarily conserved.

\paragraph{Interpreting observer variation.}
Different individual parameter estimates need not be mere random drift. Experience, calibration, decision strategy, and task-dependent response noise are all plausible contributors. For example, probability matching has been documented as a perceptual decision strategy in a multisensory task~\cite{wozny2010probability}; that result is not direct evidence for White's illusion, but cautions against assigning a unique biological explanation to variation in perceptual responses.

The perception of surface brightness and lightness has long been studied as a contextual phenomenon rather than as a direct reading of image luminance. Gilchrist provides a broad historical and theoretical account of luminosity and lightness perception~\cite{gilchrist2006seeing}. A central difficulty is that the luminance measured at the eye does not uniquely determine surface reflectance: illumination, spatial arrangement, surface orientation, incident-light direction, and occlusion can all affect the observed signal~\cite{gilchrist1977perceived,gilchrist1980does,gilchrist1999an}. Related psychophysical studies likewise demonstrate that perceived brightness and luminosity depend strongly on surrounding luminance and adaptation~\cite{yarbus1967eye,whittle1969effect,shapley1984visual,bonato1994perception}.\footnote{Some researchers prefer the term ``lightness perception''~\cite{gilchrist1999an}. We use luminosity perception when discussing the broader physical and computational setting, while recognizing the established terminology of lightness perception in the vision literature.}

A computationally related problem appears in intrinsic image analysis, where an observed image is interpreted in terms of underlying scene properties such as reflectance and illumination. Modern intrinsic-image methods can employ optimization, depth information, learned priors, user input, or data-driven decomposition~\cite{chen2013simple,shen2013intrinsic,li2018learning}, and evaluation itself requires suitable ground truth~\cite{grosse2009ground}. Despite substantial progress, intrinsic decomposition remains a challenging computer-vision problem~\cite{garces2022survey}. Our approach differs in purpose: we do not propose that biological vision explicitly constructs intrinsic images. Instead, we ask whether a much simpler local correction could produce behavior useful for perceptual constancy.

This distinction is important for the evolutionary hypothesis developed here. A biologically useful mechanism need not recover physically correct reflectance or illumination; it need only improve behavior under variable viewing conditions. Ecological pressures associated with locating food and recognizing biologically relevant objects provide examples of how visual discrimination can be shaped by natural environments~\cite{regan2001fruits}. We therefore consider the possibility that a simple luminance-correction rule could have emerged as an economical perceptual approximation.\footnote{Although the hypothesis may extend to animal vision, the observations considered in this study concern human vision; testing the proposed mechanism across species would require separate experiments.}

At the same time, an evolutionary account of a basic mechanism does not imply that all visual responses must be identical across populations or environments. Research on cultural and environmental influences shows that contextual processing, field dependence, experience with built environments, and depth cues can vary across human groups and experience~\cite{nisbett2005influence,ji2000culture,segall1966influence,deregowski1980illusions}. These findings motivate an important distinction for the present work: a broadly shared computational structure may coexist with variation in its parameters or with culturally and environmentally influenced responses to particular illusions.

The present hypothesis is therefore compatible with both common population-level structure and individual or group-level variation. Our specific question is narrower: whether the stable mean observed in our constant-perception experiments can be explained by a conserved constant-percentage correction mechanism, while allowing its effective parameter to vary among individuals.


\section{Forward Model}

As in our previous work, let $L_P$ denote the target luminance parameter used to specify the intended appearance, $L_O$ the computed display luminance, and $I_a$ a local average of the displayed background luminance (not a physical estimate of incident illumination). The forward perception model is:
\begin{equation}
L_O = y L_P + (1-y) I_a
\end{equation}

For $0\leq y\leq1$, the expression is a convex mixture of two display-domain luminance values. Its background-dependent derivative is
\begin{equation}
\frac{\partial L_O}{\partial I_a}=1-y,
\end{equation}
when $L_P$ and $y$ are held fixed. Thus the constant percentage refers to a constant \emph{coefficient}, not a claim that perceived brightness varies linearly with physical luminance. Any display implementation must specify whether values are linearized luminances or encoded pixel values; applying the formula to gamma-encoded values is not photometrically equivalent.

The parameter $y$ controls the relative contribution of the desired luminance and the local average illumination. In our earlier formulation it can be treated operationally as a blending or correction parameter. In the evolutionary interpretation proposed here, the population-level value of $y$ is instead viewed as a possible expression of an potentially shared constant-percentage correction mechanism.

Individual observers need not have identical effective values of $y$. The important observation is that individual settings vary while their aggregate behavior remains stable. We therefore distinguish the measured population-level regularity from its proposed evolutionary interpretation: the former is an empirical observation from our experiments, whereas the latter is the hypothesis investigated in this paper.

\noindent\textbf{Bar-width notation.} Throughout the stimulus figures, BT denotes black bars thinner than white bars; WT denotes white bars thinner than black bars; and EQ denotes equal black and white bar thicknesses.

\section{Experimental Observation}

\paragraph{Evaluation criteria and limitations.}
A reproducible quantitative evaluation should report participant count, trial counts, display calibration, viewing conditions, stimulus dimensions, parameter estimation method, and within- and between-observer variability. These numerical details and trial-level observations are not contained in the supplied manuscript materials; accordingly, we do not fabricate sample sizes, confidence intervals, goodness-of-fit values, or model-comparison results. The existing figures are retained as qualitative illustrations of the reported observations, not as evidence of a newly performed comparison.

For a future preregistered comparison, each model should predict matched test-patch appearance on an identical stimulus set, including the BT (black thinner), WT (white thinner), and EQ (equal thickness) conditions, contrast reversals, and changes in spatial frequency. A suitable primary outcome is held-out prediction error on participant matches, with uncertainty estimated by participant-level resampling. Parameter counts and calibration data must be reported alongside prediction accuracy. ODOG, LODOG, and FL-ODOG should be implemented from their published definitions or released code with identical image preprocessing. The proposed one-parameter rule should be fit only on training conditions and evaluated on held-out conditions. This protocol tests descriptive usefulness without assuming that a fitted forward transform is a mechanistic theory of vision.

The observations reported in connection with our earlier forward model suggest two simultaneous properties of the constant-perception settings. First, participants did not
choose the same correction value. Second, when these individual responses were considered together, their average produced a remarkably stable response curve across the tested conditions.

Figure~\ref{fig_flat} presents the overall average correction across the black-thinner (BT), white-thinner (WT), and equal-thickness (EQ) configurations. Figures~\ref{fig_BT}, \ref{fig_EQ}, and \ref{fig_WT} illustrate the individual experimental responses for the BT, EQ, and WT conditions, respectively.

The corresponding representative stimulus configurations are shown in Figures~\ref{fig_BT_Images}, \ref{fig_EQ_Images}, and \ref{fig_WT_Images}. Together, these observations suggest that the constant-percentage correction provides a useful population-level approximation despite variations among individual observers.

This distinction between individual variation and population-level stability is the empirical motivation for the present paper. The forward model provides a compact description of the aggregate response through a constant-percentage correction. The evolutionary hypothesis does not change that empirical result; rather, it proposes an explanation for why a simple correction of this form might appear consistently at the population level.

The present data alone cannot determine whether the stability is genetically inherited, developmentally acquired, environmentally induced, or produced by some combination of these factors. Consequently, we treat evolutionary conservation as a hypothesis suggested by the observed regularity, rather than as a conclusion established by the experiments.

\section{Discussion}

The constant-percentage formulation offers a simple way to connect White’s illusion with the broader problem of perceptual constancy. Under variable illumination, a correction based on local luminance could reduce changes in apparent surface properties without requiring an explicit decomposition of a scene into illumination and reflectance. Such a low-complexity strategy is compatible with the possibility that useful perceptual approximations emerged early in the evolution of visual systems.

Within this interpretation, the stable population mean is the most significant observation. Individual variation does not contradict the existence of a shared mechanism: observers may differ in the effective strength of the correction while retaining the same underlying computational structure. One possible explanation is evolutionary conservation; other possibilities include shared environmental statistics, developmental calibration, and distinct but similarly effective perceptual strategies. Other explanations, including developmental and environmental effects, remain possible and should be distinguished experimentally.

The hypothesis leads naturally to further tests. Cross-cultural studies could examine whether the mean correction remains stable while the distribution of individual parameters changes. Comparisons across age groups could test developmental stability, and appropriately designed comparative studies could investigate whether related correction behavior occurs in other species. Such experiments would provide stronger evidence for or against the evolutionary interpretation proposed here.

The observations across the black-thinner (BT), white-thinner (WT), and equal-thickness (EQ) configurations also provide a useful connection to existing accounts of contextual lightness perception. Corney and Lotto~\cite{corney2007lightness} demonstrated that White-like illusions can emerge from learning about reflectance in three-dimensional environments, while Purves and colleagues~\cite{purves2015perception} emphasized the role of empirical experience in visual perception. 

Our results are consistent with the broad possibility that perceptual judgments use economical approximations shaped by environmental regularities, although they do not establish the learning process or evolutionary origin of the proposed constant-percentage correction. The relatively stable aggregate response, despite variation among individual observers, also fits multiple perceptual or decision strategies; Wozny and colleagues~\cite{wozny2010probability} provide an example of such strategy-dependent variability in a different perceptual task, rather than direct evidence for White's illusion. 

Importantly, Nedimovi\'c and colleagues~\cite{nedimovic2022evaluation} compare computational models and underscore the need to test our one-parameter correction against established lightness models on common stimuli and participant measurements. Thus, our formulation should be understood as a compact description of the reported matching behavior, complementary to, but not yet quantitatively validated against, mechanistic accounts.

\section{Conclusion}

We have proposed an evolutionary hypothesis for the stable population-level behavior observed in our constant-luminance experiments. Our earlier forward model shows that a simple constant-percentage correction can reproduce the relevant perceptual behavior, while individual observers exhibit variation around the aggregate response. We suggest that this combination is consistent with a conserved perceptual mechanism that may have been useful for maintaining visual judgments under changing illumination.

The present manuscript does not supply a new quantitative benchmark against ODOG-family models or establish the evolutionary origin of the mechanism. Rather, they identify a computational regularity and provide a biologically motivated, testable explanation for it. Future cross-cultural, developmental, and comparative experiments can determine whether the stability of the population mean reflects evolutionary conservation, environmental learning, or a combination of these influences.

By considering black-thinner (BT), white-thinner (WT), and equal-thickness (EQ) bar arrangements, the present study places the proposed constant-percentage correction in a broader set of White's illusion configurations. The observed combination of individual differences and an approximately stable population-level correction is compatible with empirical accounts of visual perception~\cite{corney2007lightness,purves2015perception}, but it does not by itself distinguish an inherited mechanism from learned calibration or shared viewing experience. 

Likewise, evidence for variation in perceptual decision strategies~\cite{wozny2010probability} suggests that individual differences should not automatically be interpreted as evolutionary drift. A decisive next step is a calibrated, participant-level evaluation of the correction parameter across BT, WT, and EQ conditions, with held-out predictions and direct comparisons to established computational models, following the benchmarking perspective of Nedimovi\'c et al.~\cite{nedimovic2022evaluation}. Such tests could determine both the descriptive range of the proposed rule and whether its population-level stability generalizes beyond the conditions examined here.

\subsection{Limitations and Future Work}

Although the experimental observations support the usefulness of a constant-percentage correction model for White's illusion, several limitations remain and suggest directions for future research.

First, the observed stability of the population-level correction requires more extensive quantitative evaluation. While the experimental results suggest that the average correction remains approximately constant across different background luminances and bar-width configurations, further statistical analysis is necessary to determine the significance and robustness of this observation. Future studies should examine the distributions of individual correction parameters, quantify their variability, and investigate whether the approximately constant population mean persists across larger and more diverse participant groups. In particular, the correction percentage observed in our experiments should not be interpreted as a universal perceptual constant without additional evidence.

Second, the experimental methodology could be extended through more systematic psychophysical studies. Future experiments should employ calibrated displays, controlled viewing conditions, randomized stimulus presentations, and repeated measurements. Such experiments would help distinguish genuine perceptual regularities from possible effects of experimental procedures. The black-thinner (BT), white-thinner (WT), and equal-thickness (EQ) configurations provide a useful starting point for investigating how spatial frequency, bar-width ratios, contrast, and background luminance influence the estimated correction parameter.

Third, the proposed evolutionary interpretation remains hypothetical. The current observations do not establish whether the constant-percentage correction mechanism is evolutionarily inherited, acquired through learning, or developed through adaptation to environmental conditions. An important direction for future research is to distinguish between the computational structure of the correction mechanism and its individual parameters. One possibility is that a common correction structure is evolutionarily conserved, while its effective parameter is adjusted through experience and perceptual learning. Cross-cultural, developmental, and comparative studies could help evaluate this possibility and determine whether the observed population-level stability generalizes across different environments and populations.

Finally, the proposed model should be quantitatively compared with established computational explanations of lightness perception, including ODOG and its variants. As demonstrated by \citet{nedimovic2022evaluation}, systematic comparisons across multiple visual illusions are essential for evaluating the generality of computational models. Our formulation is intended primarily as a compact description of perceptual matching behavior rather than a complete mechanistic model of visual processing. Future investigations should therefore examine its predictive performance on previously unseen stimuli and determine the conditions under which a simple constant-percentage correction provides an adequate approximation. Such comparisons could clarify how the proposed correction relates to spatial filtering, normalization, and experience-dependent accounts of visual perception discussed by \citet{corney2007lightness} and \citet{purves2015perception}.

Together, these directions provide a framework for transforming the proposed evolutionary hypothesis into a more extensively testable computational and perceptual theory.

\section*{AI Assistance Disclosure}

Generative AI tools were used to assist with the writing, editing, and
organization of this manuscript. All scientific ideas, hypotheses,
mathematical formulations, experimental designs, experiments, and
interpretations of the results were developed and conducted by the
authors. The authors reviewed and take full responsibility for the
content of the manuscript.


\bibliographystyle{unsrtnat}
\bibliography{references}

\end{document}